
\input harvmac
\Title{HUTP-91/A062}{Exact Results for Supersymmetric Sigma Models}
\vglue 1cm
\centerline{S. Cecotti}
\centerline{International School for Advanced Studies, SISSA-ISAS}
\centerline{Trieste and I.N.F.N., sez. di Trieste, Trieste, Italy}
\vglue 1cm
\centerline{{Cumrun Vafa}}
\centerline{Lyman Laboratory of Physics}
\centerline{Harvard University}
\centerline{Cambridge, Ma 02138, USA}
\vglue 2cm

We show that the metric and Berry's curvature
for the ground states of $N=2$ supersymmetric sigma
models can be computed exactly as one varies the Kahler structure.  For the
case
of $CP^n$ these are related to special solutions of affine toda equations.
This allows us to extract exact results (including exact instanton
corrections).
We find that the ground state metric is non-singular as the size of the
manifold shrinks to zero thus suggesting that 2d QFT makes sense even
beyond zero radius.  In other words it seems that
manifolds with zero size are non-singular as target spaces
for string theory (even when they are not conformal).
 The cases of $CP^1$ and $CP^2$ are discussed in more
detail.

\Date{11/91}

In this letter we discuss some exact
non-perturbative results for $N=2$ supersymmetric sigma
models by applying the results developed in \ref\cv{
S. Cecotti and C. Vafa, {\it Topological anti-Topological Fusion},
1991 preprint, HUTP-91/A031; SISSA-69/91/EP.}.
Let us briefly recall the main results in that paper.
Consider a two dimensional QFT with $N=2$ supersymmetry.  Let
$Q^+_\alpha$ and $Q^-_\alpha$ label the two supersymmetry charges
($\alpha$ denotes the chirality).  The ground states $|a\rangle $
of the supersymmetric QFT are characterized by
$$Q^{\pm}_\alpha |a\rangle =0.$$
(where we take the space to be a circle of length $1$ with periodic
boundary conditions).  Operators $\phi_i$ which commute with $Q^+$
$$[Q^+_\alpha ,\phi_i]=0$$
are called chiral.  The $CPT$ conjugate operators $\bar \phi_i$
commute with $Q^-$ and are called anti-chiral.  There is a one to one
correspondence between the ground states and the chiral operators
(as can be seen by applying a chiral field to a canonical ground state
represented by $|1\rangle $,
which is always uniquely definable \cv , and
projecting it to the canonical ground state subsector).  We can thus label
the ground states by the labels of chiral fields $|i\rangle $.
$$\phi_i |1\rangle =|i\rangle +Q^+|\psi \rangle$$
Similarly we can use the anti-chiral operators to label the same
ground states, but in a different basis $|\bar j \rangle$ (which
is the conjugate to $\langle j|$).
The chiral fields form a commutative associative ring
$$\phi_i \phi_j=C_{ij}^k \phi_k +[Q^+,\Lambda ]$$
We identify the action of $\phi_i$ on the chiral fields with the matrix
$C_i=C_{ij}^k$.  This matrix also represent
the action of $\phi_i$ on the ground states:
$$\phi_i |j\rangle =C_{ij}^k|k\rangle +Q^+|\psi \rangle$$
We can use the top components of $\phi_i$ and its conjugate
to perturb the action in a supersymmetric way
$$\delta S= \int d^2\theta d^2z\  \delta t^i \ \phi_i+ c.c.$$
As we perturb the action by changing $t^i$ the ground states change.
Introduce the connections
$$A_i=\langle \bar j | \partial_i |k\rangle $$
and its conjugate, which are defined as a
function of $t_i,\bar t_i$ and act on the space of ground states.
This connection `measures' the way the ground state subsector
varies in the full Hilbert space as we change the couplings
\ref\wil{F. Wilczek and A. Zee, Phys. Rev. Lett. 52 (1984) 2111.}.
It is easy to see that as we change the ground state
basis $A_i$ transforms as a gauge connection.
Consider the covariant derivatives
$$D_i=\partial_i -A_i \ \ \bar D_i=\bar \partial_i -\bar A_i $$
Let $g$ denote the Hermitian metric in the ground state subsector
labelled by the chiral fields:
$$g_{i\bar j}=\langle \bar j |i \rangle $$
It is natural to define in addition a `topological' metric $\eta$
which is symmetric and given by
$$\eta_{ij} =\langle i|j\rangle $$
There is a relation between $g$ and $\eta$;
The relation follows from the fact that $|{\bar i}\rangle$
is CPT conjugate of $|i\rangle$ and we get
\eqn\rea{g^{-1}\eta (g^{-1} \eta )^*=1}

By the above definition it simply follows that $D_ig=\bar D_i g=0$.
The main result of \cv\ is to derive a set of differential
equation which $g$ and $A$ satisfy as functions of couplings
$t_i,{\bar t_j}$.  The equations which we will mainly use
give us the curvature of $A$ (which is the generalization
of the Berry's curvature to degenerate ground states \wil )
 in terms of the commutator of the chiral and
anti-chiral rings:
$$[D_i,\bar D_j]=-[C_i, \bar C_j]$$
$$[D_i,D_j]=[\bar D_i ,\bar D_j]=0$$
Choosing a holomorphic gauge ($\bar A_i =0$)
the first equation can be rewritten (using covariant
constancy of $g$) as
\eqn\curv{{\bar \partial_j} (g{ \partial_i} g^{-1})=[C_i,g(C_j)^\dagger
g^{-1}]}
This is the main equation we shall use in this paper to solve for $g$.

We now wish to apply this formalism to supersymmetric sigma model
on a Kahler manifold $M$. In such a case the ground states, and thus
the chiral fields are in one to one correspondence with cohomology elements
of $M$ \ref\wita{E. Witten, Commun. Math. Phys. 118 (1988) 411 ;
Nucl. Phys. B340 (1990) 281.}.
  Moreover $\eta$ may be identified with the intersection of cohomology
elements. The chiral rings, ignoring instanton corrections
may be identified with the cohomology ring
of $M$.  It turns out to be possible to find the
exact modification to the chiral ring due to instanton
corrections \wita .
We then simply use \curv\ (and the constraint
\rea ) to solve for $g$ (and the ground state curvature).

Instead of being general let us consider the case of supersymmetric
$CP^{n-1}$ sigma models which has been studied extensively in the literature.
The classical cohomology ring is generated by a single element $x$  (of
dimension $(1,1)$) represented as a bilinear in fermions with the relation
$$x^n=0$$
which is modified by instanton corrections simply
to \wita \ref\ki{K. Intriligator, {\it Fusion Residues}, Harvard preprint,
HUTP-91/A041.}
\eqn\ring{x^n=\beta}
(reflecting the fact that $2n$ fermions can absorb the zero
modes in the presence of instanton)
where $-\rm log\ \beta$ is the action for a holomorphic instanton
(which wraps the sphere once around the non-trivial two cycle of $CP^{n-1}$).
Note that $\beta$ need not be real as we can add a topological term
to the action which gives a phase modification only in the presence of
instantons.  Since the real part of the action is positive we have $|\beta
|\leq 1$.  The topological metric $\eta_{ij} =
\langle x^i x^j\rangle=\delta_{i+j,n-1}$
in the natural basis corresponding to $\int x^{n-1} =1$.
Since $x$ represents the Kahler class, the top component of it
is related to the action itself.  So we can write the action as
\eqn\action{S=-{\rm log} \ \beta \int d^2zd^2\theta \ x +c.c.}
Variation of $\beta $ brings down $-x\over \beta$, which is represented
in the basis of monomials $x^i$ by the matrix
$$C_\beta ={-1\over \beta}\left(\matrix{0&1&0&\ldots&0&0\cr
0&0&1&\ldots &0&0\cr \ldots&\ldots &\ldots&\ldots&\ldots&\ldots \cr
\ldots&\ldots &\ldots&\ldots&\ldots&\ldots \cr 0&0&0&\ldots&0&1 \cr
\beta&0 &0 &\ldots&0 &0\cr}\right)$$
Note that $x^i$ has (left,right) fermion number $(i,i)$.  But fermion
number is violated by multiples of $2n$
units due to instanton effects.  So we still
have a $Z_{2n}$ conservation of chiral fermion number which in particular
implies that $g$
$$g_{i\bar j }=\langle {\bar x^j}| x^i \rangle $$
is diagonal.  Define
$$q_i=\rm log \ g_{i\bar i}-{2i-n+1\over 2n}\rm log |\beta|^2$$
$$z=n\ \beta^{1\over n}$$
Then equation \curv\ becomes
$$\partial_z \partial_{\bar z} q_i+ e^{(q_{i+1}-q_i)}-e^{(q_i-q_{i-1})}=0$$
with $q_{n}$ defined to be the same as $q_0$.  This equation is the familiar
affine $\hat A_{n-1}$ Toda equation.
Using \rea\ and the form of $\eta$ we learn that $q_i+q_{n-1-i}=0$ which
reduces the above equations to $\hat C_m$ ( $\hat BC_m$)
Toda equation where $n=2m$ ($n=2m+1$).

The metric $g$ is a function of $|\beta |^2$.  This is due to the fact
that we need to have equal number of instantons and anti-instantons to
get a non-zero contribution to $g$ (otherwise fermion zero modes will
kill the contribution).  So the above equations become 1 dimensional.  In
other words we are looking for radially symmetric solutions of affine Toda
equations.  The only additional ingredient we have to provide to completely
solve the above equations is the boundary condition.  We do this near $\beta
\sim 0$, where the radius of $CP^{n-1}$ goes to infinity, and we can use
semiclassical arguments to find the norm of the states by
representing them as harmonic forms. Namely
\eqn\semc{\langle {\bar x^r}|x^r \rangle =2^{n-1-2r}\int x^r\wedge *x^r =
{r!\over (n-1-r)!}(-2\rm log
|\beta |)^{n-1-2r}}
(here we used that
the K\"ahler form $k=-{\rm log}|\beta| x$ and that
$*k^r=(r!/(n-1-r)!) k^{n-1-r}$).
However, this ignores the loop corrections; it turns out that
the solutions to the toda equation themselves know about loop corrections!
As is  well known the only loop correction to the Kahler {\it class}
is the one loop result \ref\alg{L. Alvarez--Gaum\'e and
P. Ginsparg, Commun. Math. Phys. 102 (1985) 311.}\foot{It is conceivable
that there may be loop corrections to the
composite operators corresponding to $k^r$.}
 which makes the coupling dynamical.
The RG flow in this case predicts that
\eqn\rgf{-\rm log \beta =c_1 \rm log \mu }
where $c_1=n$ is the first chern class for $CP^{n-1}$ and $\mu$ defines the
RG scale.
 This is reflected in our formalism by the fact that
if the size of the one dimensional circle is $L$ instead of
$1$ the dimensionless quantity appearing in the solution changes
from $\beta^{1/n}\rightarrow L \beta^{1/n}$, in other words $\beta^{1/n}$
has secretly the dimensions of mass (which could also be inferred from the
fact that $x^n=\beta $ and $x$ is a fermion bilinear);
by introducing a mass scale $\mu$
we can thus write the solution as a function of
$L\mu \beta_0^{1/n}$ where $\beta_0$ is again dimensionless.
This indeed tells us that $\beta_0$ flows with RG scale according to \rgf .
However it turns out that
the one loop computation has a finite left-over
piece which gives a finite quantum correction to the effective
coupling $\beta$; in the minimal subtraction scheme we find the correction
to be $-{\rm log} \beta \rightarrow -{\rm log}\beta -c_1 \gamma $
where $\gamma$ is the Euler's constant.
So a more accurate semiclassical
computation which takes into account loop corrections should
replace \semc\ by
$$\langle {\bar x^r}|x^r\rangle ={r!\over (n-1-k)!}[2
({\rm -log}|\beta |-n\cdot \gamma )]^{n-1-2r}$$

The equations we get for the case of $CP^1$ and $CP^2$ turn out to have
been studied extensively \ref\wu{B. M. McCoy, C.A. Tracy and T.T. Wu,
J. Math. Phys. 18 (1977) 1058.\semi A.R. Its and V. Yu.
Novokshenov, {\it The isomonodromic Deformation Method in the
Theory of Painlev\'e Equations}, Lecture Notes in Mathematics 1191,
Springer--Verlag, Berlin 1986.}\ref\kit{A.V. Kitaev, {\it The Method of
 Isomonodromy Deformations for Degenerate Third Painleve Equation}, published
in
{\it Questions of Quantum Field Theory and Statistical Physics} 8 (Zap. Nauch
Semin. LOMI v.161), V.N. Popov and P.P Kulish (ed.) Nauka (Leningrad). }.
After
imposing the reductions discussed above the
equation for $CP^1$ case becomes the special case
of Painleve $PIII$ equation. Then the only consistent solutions which
have no pole as a function of $\beta $ which have a logarithmic $\beta$
dependence can only be of the form (after changing back to physical
variables) \wu\ as $\beta \rightarrow 0$,
$$\langle \bar x |x\rangle ={1\over 2(\rm -log |\beta |-2 \gamma )}
(1+O(|\beta |^2\rm
log^2|\beta |)).$$
 Needless to say we can compute order by order
the instanton contributions by solving the differential equation.
Indeed,
for small $\beta$, $\langle \bar x|x\rangle^{-1}$ has an expansion of the form
\ref\ising{T.T. Wu, B.M. McCoy, C.A. Tracy and E. Barouch, Phys. Rev.
 B13 (1976) 316.}
$$\langle \bar x|x\rangle^{-1}=\sum_{n=0}^{\infty} |\beta|^{2n}p_n,$$
where $p_n$ is a polynomial in $[-{\rm log}|\beta|-2 \gamma ]$ of degree
$2n+1$.
The coefficients of these polynomials can be computed recursively from the
differential equation\foot{For $n\leq 3$ these coefficients are listed in
Ref.\ising .}. The corrections clearly reflect the contribution of
$n$ instanton anti-instanton pairs
to the metric (as they have a prefactor of $|\beta|^{2n}$).  It would
be interesting to compare the $p_n$ from perturbations about
the instanton backgrounds.
It is also very satisfactory that the form of the loop corrections are
{\it predicted} to be that given by the one loop term (including the
Euler's constant) simply by requiring non-singularity of the solution
as a function of $\beta$.  In fact the solution can be continued
analytically even passed $\beta =1$ which corresponds to zero radius
on the sphere to $\beta \geq 1$ which has no obvious relation to sigma
model on sphere.  We should have expected that we will not encounter
any singularities because according to renormalization group flow
computation \rgf\ in {\it finite} RG time (i.e. finite mass scale) we come
to have a zero radius and if the theory is sensible at all we should
be able to choose any mass scale which would correspond to passing
through zero radius!  In fact the asymptotic structure to the metric
has also been worked out as $\beta \rightarrow \infty$,
and one finds that
$${\langle \bar x |x\rangle \over \langle 0|0\rangle}=|\beta |(1-{1
\over (\pi^2 |\beta |)^{1/4}}\rm exp [-8|\beta |^{1/2}])+...$$
The interpretation of this is as follows:
the ring $x^2=\beta$ suggests that for large $\beta$ the field
configuration is dominated near $x=\pm \sqrt \beta$, which explains the
leading term in the above asymptotic behavior.  The subleading
exponential term
suggests that there is tunneling between these configurations by a
soliton with mass $8|\beta |^{1/2}$.  The behavior
of this theory is very similar to that of the Landau-Ginzburg theory
which has the same ring, namely $W=x^3/\beta -3x$.  The difference being
that the behavior as $\beta \rightarrow 0$ is different between the two
and thus we get a different solution of PIII equation.  The $\beta
\rightarrow \infty$ behavior for the LG theory is the same as the above,
with the difference that the coefficient in front of the exponential term
is smaller by a factor of 2.  In \cv\ this coefficient was (heuristically)
related to the number of particles, and so this suggests that in the $CP^1$
model we have twice as many particles as in the LG theory, which is indeed
the case, since $CP^1$ is believed to have \ref\mas{
E. Abdalla, M. Forger and A. Lima Santos, Nucl. Phys. B256 (1985) 145\semi
E. Abdalla and A. Lima Santos, Phys. Rev. D29 (1984) 1851\semi
V. Kurak and R. Koberle, Phys. Rev. D36 (1987) 627.}\
one doublet of $SU(2)$ whereas LG theory has only one particle.

This story can be repeated exactly as before for $CP^2$, where the
relevant equation has been studied in \kit\ with the result that
(with $s=3$,$g_1,g_2=0,\ g_3=1$ in his notation) we get a non-singular
solution, where the boundary condition is again predicted to be
the same as \semc\ with the coefficient in front of $\gamma $ being $n=3$
the first chern class of $CP^2$.  Again the asymptotic behavior is worked out
and one finds a similar behavior, this time predicting the existence
of particle of mass $6{\sqrt 3}|\beta |^{1/3}$.  Moreover the strength of the
soliton correction is three times bigger than the LG theory $W=x^4/\beta
-4x$, suggesting that we have three times as many particle.  Which is indeed
the case, as $CP^2$ is believed to have six particles (3 and $\bar 3$)
\mas\ where the LG theory is believed to have only 2 (which
has been recently confirmed in \ref\recen{
P.Fendley and K.Intriligator, in preparation.}).

Needless to say we believe that similar behavior will work for general
$CP^n$, the difference being that there the explicit solution
to (reduced) affine toda has not
been worked out.  However using physics we can predict what would be
a non-singular solution to the corresponding Toda equations.  Moreover
the result in the appendix B of \cv\ shows that from the $\beta \rightarrow
\infty$ we can read off the spectrum of masses of particles and this turns
out to be $4n\ {\rm Sin} (\pi r/n)|\beta|^{1/n}
$ as $r$ runs from $1$ to $n-1$ in agreement with \mas .
One should in principle be able to read off degeneracies as well
by a careful study of the strength of the soliton corrections to determine
the degeneracy of particles as we did for $CP^1$ and $CP^2$.
We can repeat this game for any manifold for which we know the
quantum cohomology ring.  This is in fact conjectured for
Grassmanians in \ref\cum{C. Vafa, {\it Proceedings of the Mirror Symmetry
Conference, MSRI, 1991}, Harvard preprint HUTP-91/A059.} (see also \ki ); a
 simple extension should work for all Hermitian symmetric
spaces.  At any rate we can for example read off the masses for the
Grassmanians.

Probably the most important aspect of this work is that
it suggests we can go beyond zero radii for supersymmetrical sigma
models thus hinting that with or without conformal symmetry
sigma models somehow `resolve' singularities of classical geometries.
This was connected with the fact that in finite RG time we flow to
the singular geometry and completion of the moduli space {\it requires}
having something beyond the singular point which is what seems to happen.
It would be interesting to unravel the geometry beyond zero radius.
In some ways this is related to the same phenomenon for Calabi-Yau
manifolds where the boundaries of moduli space are in some cases
a {\it finite} distance away from any point, and thus completion of
string theory suggests that we can go beyond them, as has been suggested
in \ref\hub{P. Candelas, P.S. Green, and T. Hubsch, Phys. Rev. Lett.
62 (1989) 1956.}.  Some aspects of our computations are similar
to the recent one of Candelas et. al.
\ref\cand{P. Candelas, X.C. de la Ossa, P.S. Green and L. Parkes,
Nucl. Phys. B359 (1991) 21.}\ in which
they computed instanton correction to the ground state metric
(Zamolodchikov metric) on 3-fold quintic.  In fact the equation they
obtain can be rephrased as {\it standard} $ A_3$ Toda \cv\
and so is similar to the off-critical equation which happens to be {\it affine}
Toda (for example for $CP^3$).

We would like to thank E. Abdalla for discussions.  The research of
C.V. was supported in part by Packard Foundation
and NSF grants PHY-89-57162 and PHY-87-14654.

\listrefs

\end